# High-performance two-dimensional p-type transistors based on GaSe layers: an ab-initio study

*Agnieszka Kuc\*, Teresa Cusati, Elias Dib, Augusto F. Oliveira, Alessandro Fortunelli,*

*Giuseppe Iannaccone, Thomas Heine and Gianluca Fiori*

Dr. A. Kuc,[1,2] Dr. T. Cusati,[3] Dr. E. Dib,[3] Dr. A. F. Oliveira,[1,2] Dr. A. Fortunelli,[4] Prof. G. Iannaccone,[3] Prof. T. Heine,[1,2] and Prof. G. Fiori[3]

[1]Wilhelm-Ostwald-Institut für Physikalische und Theoretische Chemie, University of Leipzig, Linnéstr. 2, 04103 Leipzig, Germany
[2]Department of Physics and Earth Science, Jacobs University Bremen, Campus Ring 1, 28759 Bremen, Germany
[3]Dipartimento di Ingegneria dell'Informazione, Universit`a di Pisa, Via Caruso 16, 56122 Pisa, Italia
[4]CNR-ICCOM, Istituto di Chimica dei Composti Organometallici Via G. Moruzzi 1, 56124, Pisa, Italia

E-mail: agnieszka_beata.kuc@uni-leipzig.de, teresa.cusati@for.unipi.it

Two-dimensional (2D) thin materials have been extensively studied in the past ten years due to their promising transport properties and potential technological applications, especially in the field of nanoelectronics[1–3]. Graphene has many extraordinary properties[4,5], but it does not possess a band gap, needed in conventional field-effect transistors (FETs). However, a plethora of 2D crystals exists[6–8], many of which are intrinsic semiconductors that have been proved suitable for logical-device applications[2]. Group 13 chalcogenides are among those semiconducting 2D crystals and one of them, GaSe, has been suggested as a potential FET channel material[9,10] due to its peculiar band structure[11,12].

GaSe is a layered material that belongs to the III-VI family of semiconductors (13-16, in the IUPAC nomenclature). It is characterized by high anisotropy of its chemical bonds, with strong intralayer covalent bonds and weak van der Waals interactions between the layers[13]. Each GaSe covalent layer consists of four atomic sublayers, stacked in a Se–Ga–Ga–Se sequence: the relative positions between GaSe layers define different polytypes with different crystal symmetries (see Figure 1). Although electrical properties of bulk GaSe have been studied since the 1970's[14–17], few-layer GaSe have been seldom explored for possible nanoelectronic applications[10]. Thin GaSe films exhibit very specific electronic structure, which may allow the fabrication of ultra-short-channel logical devices. The valence band is nearly flat in the central region of the Brillouin zone, resulting in very heavy holes and high density of states with sharp peaks appearing similar to van Hove singularities at the Fermi level. Such a peculiar property could be exploited to solve one of the biggest problems that affect semiconductor industry, i.e., scaling down to the sub-5 nm channel length regime. Indeed, at such dimensions, intra-band tunneling is detrimental for obtaining small currents in the OFF state, which in principle could be achieved by utilizing materials with large effective masses[18]. From this point of view, layered GaSe represents a promising material. The particular shape of its valence band, commonly referred to as *Mexican hat*[11], suggests that a very low intra-band tunneling could be achieved, yielding low static power-dissipation and large drive-currents in the ON regime of transistors.

In this work, we assess the performance of short-channel p-type GaSe-based devices through a multiscale approach, investigating the electronic structure of GaSe layered materials using DFT simulations, which serve as basis for parameterization of a tight-binding Hamiltonian expressed on a basis set of maximally-localized Wannier functions (MLWFs) to be employed in non-equilibrium Green's function (NEGF) device simulations (see *Computational Details* in SI for details on the methodology). We show that GaSe monolayers are indeed suitable for electronic p-type devices with channel length as short as 3 nm.

We have studied electronic properties of mono- (1L), bi- (2L), tri- (3L), tetra-layers (4L), and bulk structures of β ($P6_3/mmc$ space group) and ε ($P\bar{6}m2$ space group) GaSe polytypes (see Figure 1). The latter polytype is only 10 meV per unit cell (1.25 meV per atom) more stable than the β. The electronic properties of the GaSe-1L, calculated with the PBE-D3(BJ) density functional, are summarized in Figure 2. GaSe-1L is a semiconductor with indirect band gap (Δ) of about 2 eV and valence band edge shaped as a *Mexican hat* (Figures 2a and 2c), in a good agreement with the previous reports[8]. The optical band gap situated at the Γ point is only 50 meV larger than Δ. The top valence band is nearly dispersionless close to the Γ point due to the *Mexican-hat* shape of the valence band edge, which leads to sharp peaks in the density-of-states (DOS), similar in the shape as van-Hove singularities near the Fermi level[11,19–21] (typical of 1D materials) (Figure 2b). A closer look at the band structure and DOS near the Fermi level (Figure 2c and 2d) reveals a small spin-orbit splitting of about 11 meV at the Λ point, due to the lack of inversion symmetry.

The energy-dispersion color-map plot depicted in Figure 2e shows strong anisotropy of the valence band, which eventually leads to anisotropy of the effective masses along the $k_x$ and $k_y$ directions. Calculated effective masses of electrons ($m_e^*$), holes ($m_h^*$), and DOS ($m_{DOS}^*$) at the edges of the band structure are summarized in Table 1 and Table S1 in the Supporting Information (SI). The holes are very heavy (ranging from few up to tens $m_0$, depending on the $k$-space direction), while electrons are very light, with $m_e^* \leq 0.15 m_0$.

At the Γ point, the highest occupied crystal orbital (HOCO) and the lowest unoccupied crystal orbital (LUCO) of GaSe-1L are shown in Figure 2f. HOCO is formed by Se $p_z$ atomic orbitals and Ga–Ga σ molecular orbitals (formed between Ga $s$ orbitals). Se $p_x$ and $p_y$ orbitals are separated from the $p_z$ orbitals at the Γ point by about 250 meV and form the HOCO-1. The bottom of the conduction band consists of Ga and Se *s* orbitals, which form σ Ga–Se bonds. It should be noted that the results discussed so far have been obtained with the PBE-D3(BJ) density functional. If the Tran-Blaha modified Becke-Johnson potential (TB-mBJ)[22] is used instead, the electronic band gap in GaSe-1L increases by ca. 1 eV, suggesting that excitonic effects might influence the electronic properties. However, if more than one layer is involved, this energy difference reduces by one order of magnitude (ca. 100 meV). Moreover, the effect on the band dispersion is negligible and, thus, our conclusions regarding the effective masses and transport properties should remain valid.

The electronic properties of β- and ε-GaSe multilayered structures with respect to the number of layers are shown in Figures S1 and S2 in SI. Increasing the number of GaSe layers leads to a decrease of the fundamental band gap. Such a trend is similar to what has already been observed in some transition-metal dichalcogenides (TMDs), e.g., $MoS_2$.[23–25]. In the case of ε-GaSe, the difference between optical and electronic band gaps is much smaller with respect to the one observed in β-GaSe. In addition, in the ε-GaSe case, the *Mexican hat* gradually turns into a parabola, with the maximum located at the Γ point; consequently, the optical and fundamental band gaps converge to the same value, eventually leading to a direct-gap semiconductor. Our results agree well with the theoretical and experimental works of Rybkovskiy et al.[12] and Zólyomi et al.[11], who have studied mono- (1L), bi- (2L), tri- (3L), tetra- (4L) layer and bulk GaSe systems. The authors observed a dependence of the band gap with the number of GaSe layers, being larger in the thinner films.

The DOS projections in Figures S1 and S2 in SI show that each individual layer contributes with one quasi-singularity in the electronic structure; as the number of layers is increased, the DOS at the Fermi level consistently decreases (see Figure S3): at the bulk limit, the DOS projection assumes typical 3D characteristics (i.e., proportional to $\sqrt{E}$).

The crystal orbitals in β-GaSe-3L and ε-GaSe-3L are very similar to the orbitals of the monolayer (see Figures S4 and S5 in SI, respectively). It is worth mentioning that in both cases, the HOCO is composed of atomic orbitals delocalized over all three GaSe layers, whereas HOCO-1 is restricted to the central layer. In the case of β-GaSe-3L also LUCO is localized in the central GaSe layer.

The electronic structure of GaSe-1L is very sensitive to its geometry, since very small mechanical deformations lead to pronounced changes in the band structure as shown in Figure S6 in SI. The band gap decreases when a biaxial strain is applied, whereas the opposite effect is observed when the structure is compressed. Similar results have been recently reported by Huang et al.[26].

The structural deformations also modify the $p$ orbital splitting close to the Fermi level at the Γ point, namely it increases under strain and decreases under compression. On the other hand, the spin splitting at the Λ point reduces under strain and increases under compression. In addition, a second spin splitting appears at the Π point under deformation. A similar behavior is observed for the GaSe bulk (see Figure S7 in SI).

Moreover, small compressive strain results in flatter valence bands, meaning that the holes become heavier, further improving the transport properties for short channels. This effect is favorable up to about 1-1.1% of strain. Beyond this compression, the valence band changes to the parabolic shape at the Γ point.

Leveraging the GaSe valence band properties calculated from first-principles (i.e., large hole effective mass), we have performed electronic transport simulations of p-type field-effect transistors (FETs) based on GaSe thin films, as described in *Computational Details* in SI. In particular, we have focused on devices in the sub-5nm range, which represents a technologically relevant case for the semiconductor industry, especially while exploiting different materials for n- and p-type devices as suggested by [27]. We have considered devices with gate length ($L_G$) ranging from 3 nm to 5 nm, with an oxide thickness ($t_{ox}$) of 0.5 nm and a supply voltage ($V_{DS}$) of 0.54 V. In order to control short-channel effects, we have modelled a double-gate transistor with p-doped source and drain reservoirs, using $SiO_2$ as gate dielectric (Figure 3a). We have considered GaSe-1L (Figure 3) and β-GaSe-3L models (see Figure S8 in SI), both with large hole effective masses, to study the influence of the film thickness on the electronic transport.

The simulated transfer characteristics of GaSe-1L-based FETs with different gate-lengths are shown in Figure 3 (b and c). The sub-threshold slope (*SS*) – the inverse slope of the $I_{DS}$-$V_{GS}$ curve in the logarithmic scale in the sub-threshold regime – is a good figure of merit in order to evaluate the influence of short-channel effects. *SS* should be as close as possible to 60 mV dec$^{-1}$ (the minimum value achievable at room temperature in fermionic devices). In addition, the ON/OFF-current ratio should be larger than $10^4$, as required by the *International Technology Roadmap for Semiconductors* (ITRS)[28]. In the considered range of gate voltages, the GaSe-1L devices with 4 nm and 5 nm channels have very similar properties, with nearly ideal *SS* values (see Table 1).

Our results show that GaSe-1L-based transistors are barely affected by short-channel effects. The *Mexican-hat* shape of the valence band edge leads to very high effective masses in the central region of the Brillouin zone, yielding both low-tunneling current in the OFF regime and high drive currents in the ON regime.

Figure S9 in SI shows the transfer characteristic results for the β-GaSe-3L FETs with a body thickness of 2.1 nm. In sharp contrast with the monolayer case, large short-channel effects can be observed, especially for the 3 nm channel, resulting in figures of merit incompatible with the ITRS requirements [28].

We should point out that, although β-GaSe-3L is not a good choice for sub-5 nm devices, it still has fairly heavy holes and should provide means for potential logical devices with channel lengths larger than 5 nm.

In perspective, it can also be noted that other materials with the *Mexican-hat* shape of the valence band, which results in very heavy holes, are very good candidates for short-channel logical devices. Obvious examples are isoelectronic GaS, InS, and InSe polytypes[29–31].

In summary, we have investigated electronic structure and transport properties of multilayer GaSe with first-principles simulations. The electronic structure of GaSe polytypes shows unique features of the valence band edge, i.e., *Mexican-hat* shape, which gives rise to very heavy holes in the central region of the Brillouin zone. The very flat bands close to the Fermi level give rise to very large density of states in a shape of quasi-van Hove singularities. This peculiar and advantageous shape of the valence band is, however, strongly dependent on the number of layers and reduces for thicker films. At the same time, very light electrons are obtained at the Γ point. This interesting electronic structure is very promising for obtaining ultra-short-channel field-effect transistors in the sub-5 nm range. We report a GaSe-1L-based transistor model with sub-threshold slope (*SS*) of 92 mV dec$^{-1}$ and ON/OFF-current ratio of $1.38 \times 10^4$ for a channel length of 3 nm, complying with the requirements specified by the ITRS[28].

In the GaSe-3L case, however, the valence band edge has stronger dispersion around the Γ point, significantly reducing the electrostatic control of the channel gate.

**Supporting Information**.
Detailed description of methodology used in the present work; band structures and density of states of GaSe ε and β polytypes in equilibrium and under tensile strain; crystal orbitals of GaSe-3L systems; structural models used for transport calculations; simulated transfer characteristics of FET based on β-GaSe-3L.

**Acknowledgements**

The Deutsche Forschungsgemeinschaft (HE 3543/18-1), the European Commission (FP7-PEOPLE-2012-ITN MoWSeS, GA 317451), and European Commission under Contract No. 696656 (Project "GRAPHENE FLAGSHIP") are gratefully acknowledged. The authors thank Mr. Vladimir Bacic for his help with Quantum Espresso input files.

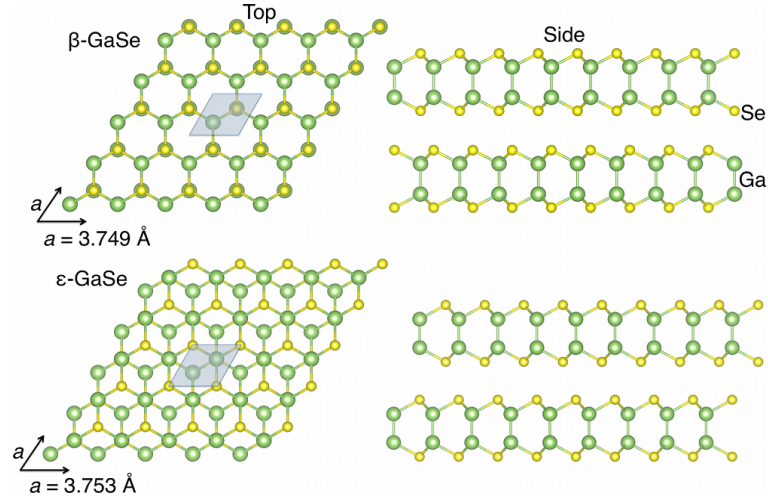

**Figure 1.** Top and side views of bulk GaSe hexagonal layered structures in β ($P6_3/mmc$) and ε ($P\bar{6}m2$) polytypes. The unit cells of both polytypes are shown in light blue and the corresponding lattice vectors are given. Ga atoms are shown in green; Se atoms are shown in yellow.

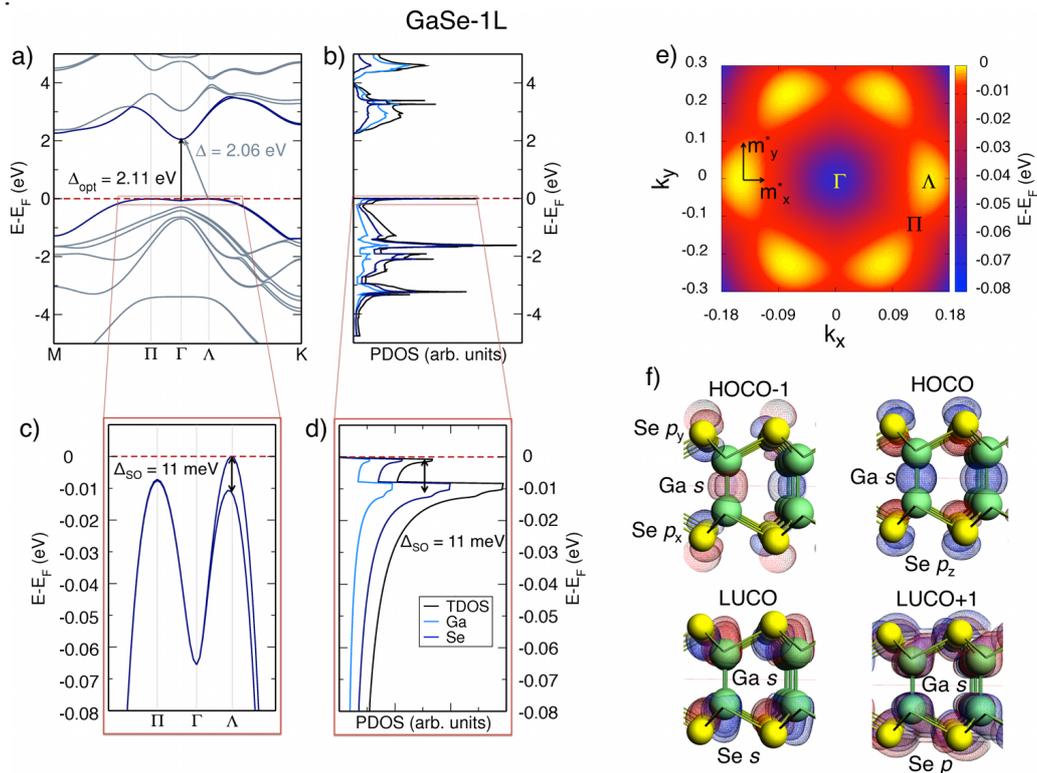

**Figure 2.** Electronic properties of GaSe-1L: a) Electronic structure and b) atom-projected density of states, and the corresponding zoom close to the Fermi level c) and d), showing ca. 11 meV spin-orbit splitting (SO). The Fermi level is indicated by the red dashed lines. Band edges are highlighted. Fundamental (Δ) and optical ($\Delta_{opt}$) band gaps are shown in grey and black, respectively. e) 3D representation of the top of the valence band close to the Γ point, showing the anisotropy of hole effective mass in $k_x$ ($m_x^*$) and $k_y$ ($m_y^*$) directions of Brillouin zone. The colored scale denotes the energy of the bands in eV. Symmetry points in the Brillouin zone are highlighted. f) Selected crystal orbitals. HOCO: highest occupied crystal orbital; LUCO: lowest unoccupied crystal orbital at the Γ point.

**Table 1.** (Left) Calculated effective masses ($m^*$) of holes ($h^+$), electrons ($e^-$), and DOS in units of electron mass ($m_0$) of GaSe-1L. The effective masses are calculated for selected $k$-points in the Brillouin zone. The band transitions (Δ) corresponding to the fundamental band gaps are also shown. $m^*_{DOS}$ is defined as $m^*_{DOS} = \sqrt{m^*_x m^*_y}$. (Right) Figures of merit ($SS$, $I_{ON}$, and $I_{ON}/I_{OFF}$) of GaSe-1L devices with different channel lengths, $L_G$.

| System | $k$ | Δ | $m^*_x$ | $m^*_y$ | $m^*_{DOS}$ | carrier | $L_G$ (nm) | SS (mV dec$^{-1}$) | $I_{ON}$ ($10^3$ A m$^{-1}$) | $I_{ON}/I_{OFF}$ ($10^4$) |
|---|---|---|---|---|---|---|---|---|---|---|
| | M | | 0.55 | 0.55 | 0.55 | $e^-$ | 3 | 92 | 1.38 | 1.38 |
| GaSe-1L | Γ | Λ→Γ | 0.15 | 0.15 | 0.15 | $e^-$ | 4 | 64 | 1.71 | 1.71 |
| | Λ | | -2.58 | -29.92 | 8.79 | $h^+$ | 5 | 62 | 1.72 | 1.72 |
| | K | | 0.63 | 0.63 | 0.63 | $e^-$ | | | | |

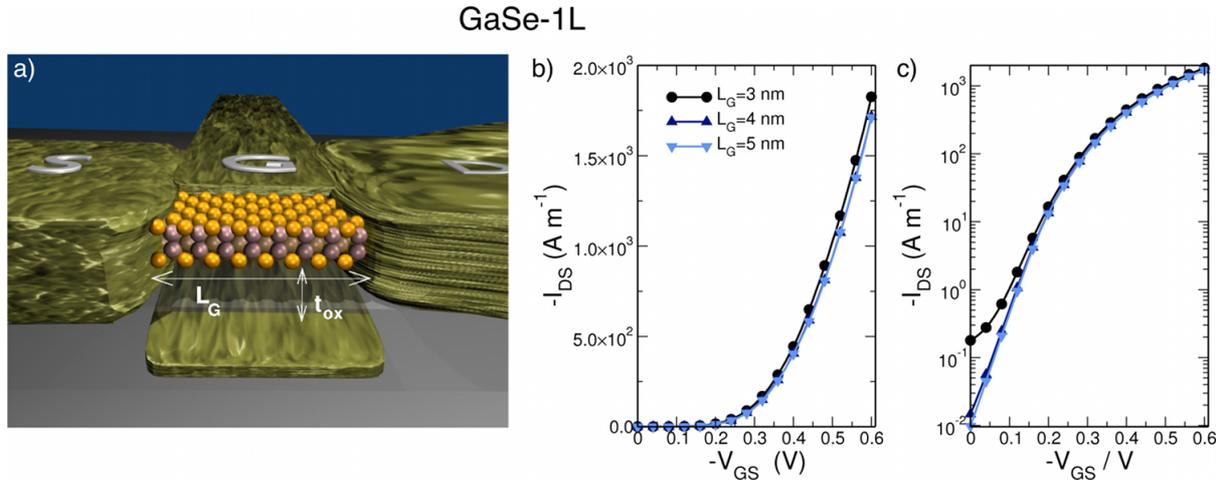

**Figure 3.** a) Schematic representation of the devices considered in this work. S: source; G: gate; D: drain. The transport channel is based on either GaSe-1L or GaSe-3L. b, c) Simulated transfer characteristics of FET based on GaSe-1L with gate lengths $L_G$ of 3, 4 and 5 nm, source-to-drain voltage $V_{DS}$ of 0.54 V, and oxide thickness $t_{ox}$ of 0.5 nm. a) linear and b) semi-logarithmic scale representation of $I_{DS}$-$V_{GS}$ curves.